\documentstyle[amsmath,amssymb,epsf,mncite]{mn}
\makeatletter\ifSFB@referee\fi\makeatother

\pubyear{1999}
\volume{666}
\pagerange{1--14}
\def\bildchen#1#2{\begin{minipage}{#1}\epsfxsize=#1\epsfbox{#2}\end{minipage}}
\def\Mpc{{\rm Mpc}}
\def\h{\ifmmode{^{\rm h}}\else{$^{\rm h}$}\fi}
\def\hMpc{\ifmmode{h^{-1}{\rm Mpc}}\else{$h^{-1}{\rm Mpc}$}\fi}
\def\hMsun{\ifmmode{h^{-1}M_\odot}\else{$h^{-1}M_\odot$}\fi}
\def\kms{\ifmmode{{\rm km}\,{\rm s}^{-1}}\else{km\,s$^-1$}\fi}
\def\bx{{\bf x}}
\def\cP{{\mathcal P}}
\def\cF{{\mathcal F}}
\def\dt{{{\rm d}t}}
\def\lV{{\overline V}}
\def\re{{\rm e}}

\begin{document}

\date{Version of 13 October 1999.  Accepted for publication in Monthly Notices.}

\title[Topology and geometry of the CfA2 redshift survey]{ 
Topology and geometry of the CfA2 redshift survey }

\newcounter{fusspilz}\setcounter{fusspilz}{0}
\def\fusspilz{{\stepcounter{fusspilz}\fnsymbol{fusspilz}}}
\author[Jens Schmalzing and Antonaldo Diaferio]{
Jens Schmalzing$^{1,2,\fusspilz}$ and
Antonaldo Diaferio$^{1,3,\fusspilz}$
\setcounter{fusspilz}{0}
\vspace*{1ex}\\
$^1$
Max--Planck--Institut f\"ur Astrophysik,
Karl--Schwarzschild--Stra{\ss}e 1,
85740 Garching, Germany.
\\
$^2$
Ludwig--Maximilians--Universit\"at,
Theresienstra{\ss}e 37,
80333 M\"unchen, Germany.
\\
$^\fusspilz$email jensen@mpa-garching.mpg.de \\
$^\fusspilz$email diaferio@mpa-garching.mpg.de \\
}

\maketitle

\begin{abstract}
We analyse the redshift space topology and geometry of the nearby
Universe by computing the Minkowski functionals of the Updated Zwicky
Catalogue (UZC).  The UZC contains the redshifts of almost 20,000
galaxies, is 96\% complete to the limiting magnitude $m_{\rm Zw}=15.5$
and includes the Center for Astrophysics (CfA) Redshift Survey (CfA2).
From the UZC we can extract volume limited samples reaching a depth of
70\hMpc\ before sparse sampling dominates.  We quantify the shape of
the large--scale galaxy distribution by deriving measures of planarity
and filamentarity from the Minkowski functionals.  The nearby Universe
shows a large degree of planarity and a small degree of filamentarity.
This quantifies the sheet--like structure of the Great Wall which
dominates the northern region (CfA2N) of the UZC.  We compare these
results with redshift space mock catalogues constructed from high
resolution $N$--body simulations of two Cold Dark Matter models with
either a decaying massive neutrino ($\tau$CDM) or a non--zero
cosmological constant ($\Lambda$CDM).  We use semi--analytic modelling
to form and evolve galaxies in these dark matter--only simulations.  We
are thus able, for the first time, to compile redshift space mock
catalogues which contain galaxies, along with their observable
properties, rather than dark matter particles alone.  In both models
the large scale galaxy distribution is less coherent than the observed
distribution, especially with regard to the large degree of planarity
of the real survey.  However, given the small volume of the region
studied, this disagreement can still be a result of cosmic variance,
as shown by the agreement between the $\Lambda$CDM model and the
southern region of CfA2.
\end{abstract}


\footnotetext{
$^3$
Present address:
Universit\`a degli Studi di Torino,
Dipartimento di Fisica Generale ``Amedeo Avogadro'',
Via Pietro Giuria 1,
I-10125 Torino,
Italy.
}

\section{Introduction}

\begin{figure*}
\bildchen{5cm}{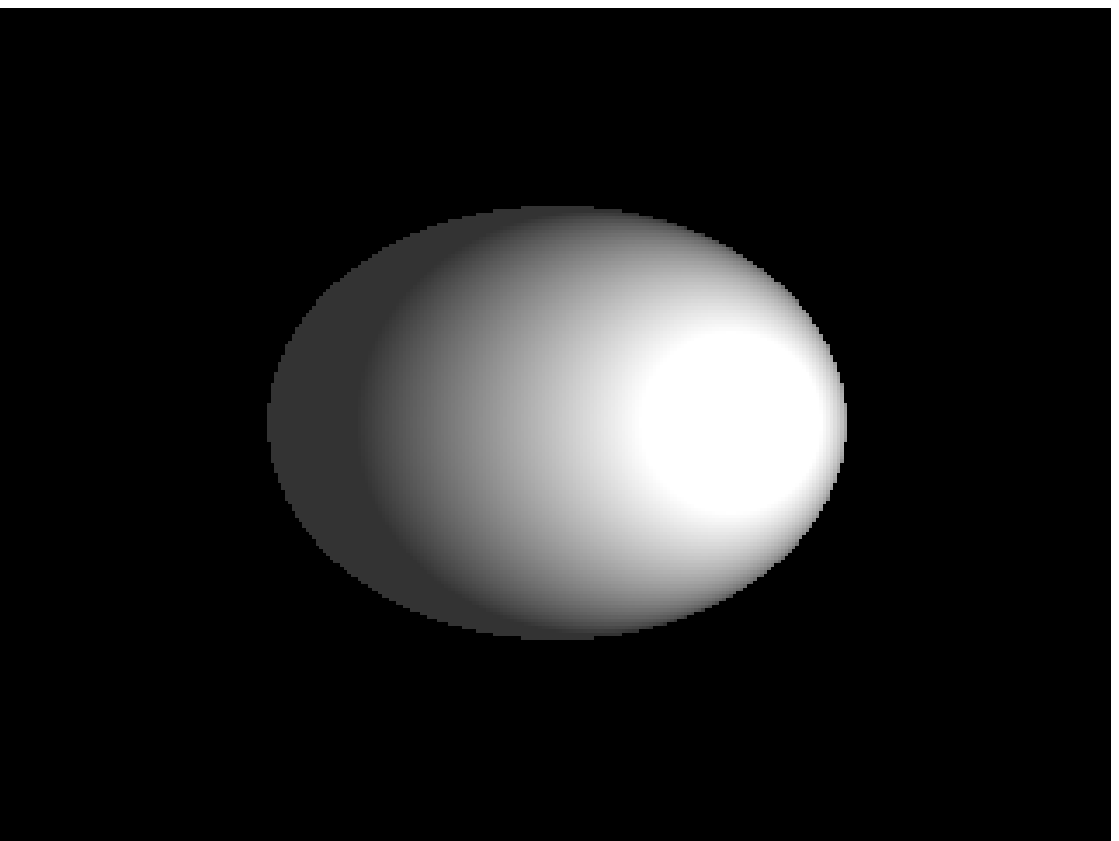}
\bildchen{5cm}{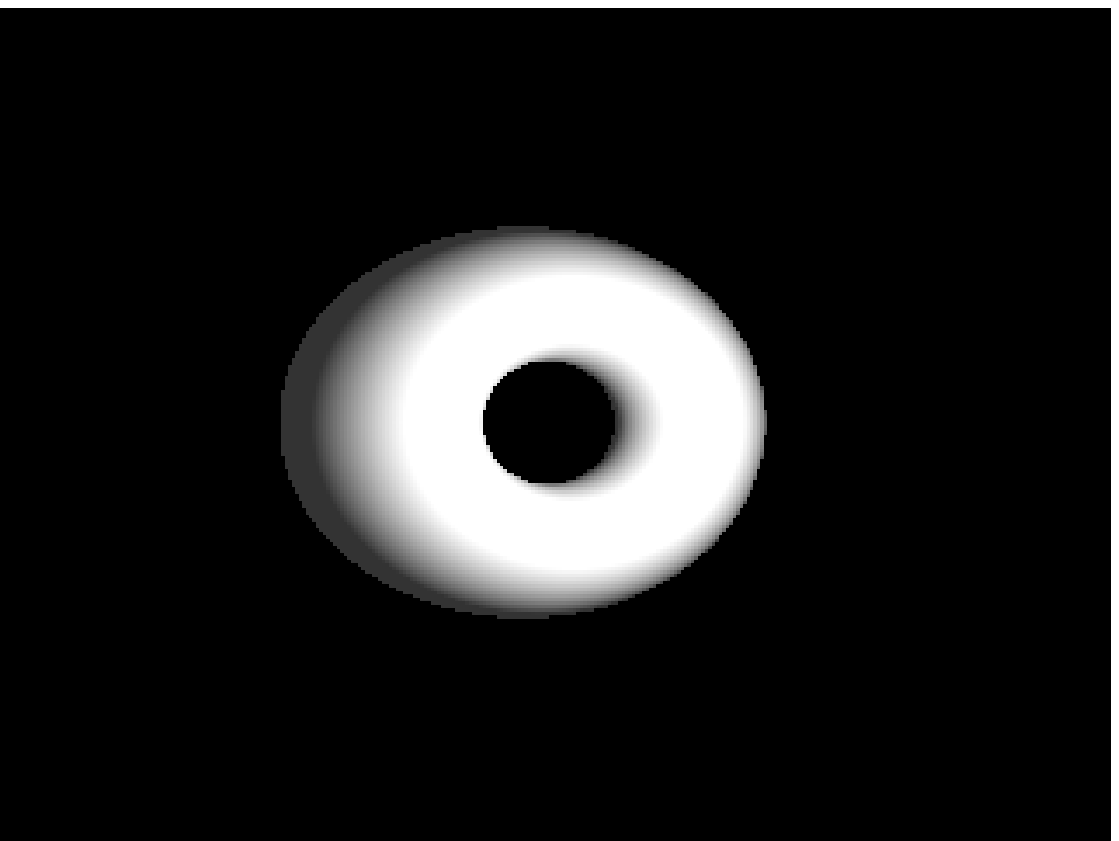}
\bildchen{5cm}{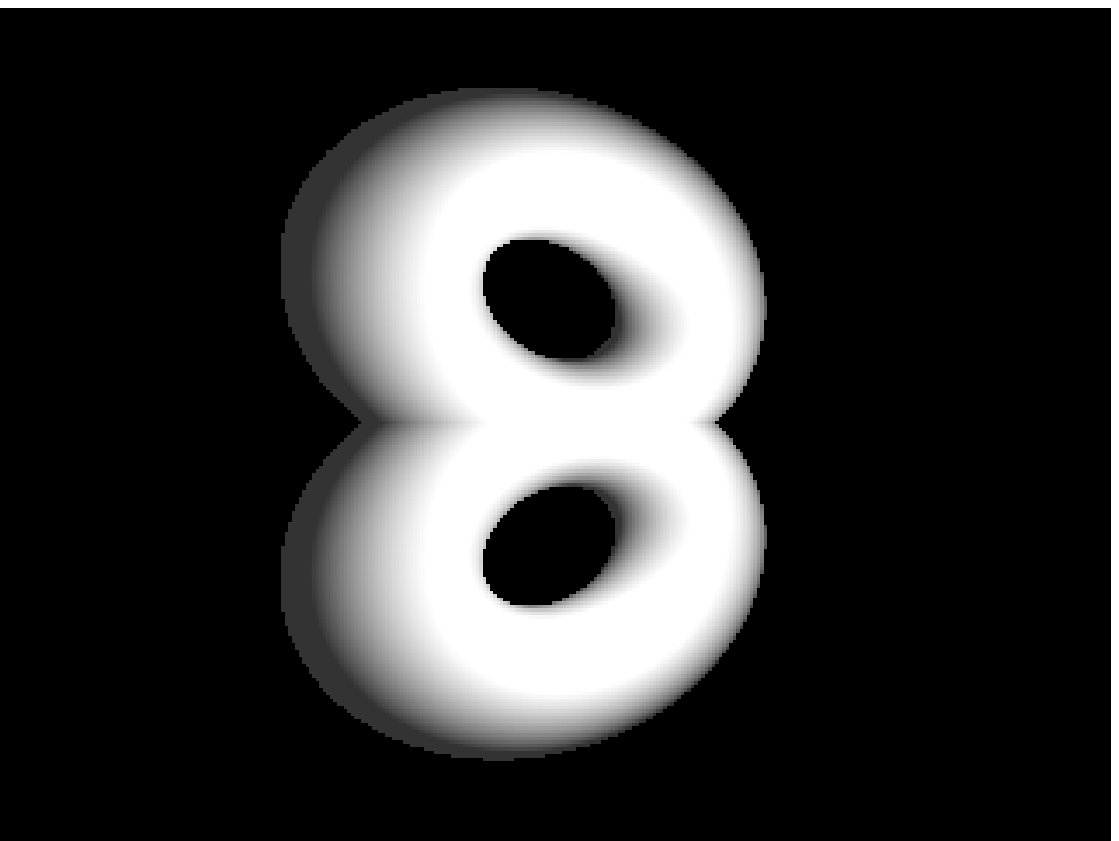}
\caption{
\label{fig:holes}
Simple structures with positive (egg, left panel), vanishing (loop,
middle panel), and negative (double loop, right panel)
Euler characteristic $V_3$.}
\end{figure*}

Standard inflationary models (e.g. {\pcite{peacock:cosmological}})
predict that the large--scale structure of the present day Universe
originates from primordial density perturbations amplified by
gravitational instability.  Under the influence of its own gravity,
the initial Gaussian random field of the density fluctuations evolves
into a strongly non--Gaussian density field.  Therefore, in order to
describe the observed galaxy distribution quantitatively, we need
statistics beyond the traditional Fourier transform pair $P(k)$ and
$\xi(r)$, the power spectrum and the two--point correlation function.
In fact, these quantities only describe a Gaussian density field in a
unique and complete way.

One possible approach to capturing higher--order features of the
galaxy distribution involves geometrical descriptors.  A number of
these statistics have been suggested to date.  Among others, recent
morphological analyses of the nearby Universe have been performed
through measuring the void probability function of the CfA2 catalogue
{\cite{vogeley:voids}}, applying percolation techniques to the
IRAS~1.2Jy catalogue {\cite{yess:percolation}} and the Las Campanas
Redshift Survey (LCRS) {\cite{shandarin:detection}}, or through various
types of shape statistics, as the moment--of--inertia method
suggested by {\scite{babul:filament}} and applied by
{\scite{dave:filament}} to the CfA1 redshift survey and by
{\scite{sathyaprakash:filaments}} to the IRAS~1.2Jy catalogue.  The
most wide--spread technique is probably the genus statistics
introduced by {\scite{gott:sponge}}, which has been used on various
IRAS catalogues
{\cite{moore:topologyqdot,protogeros:topology,canavezes:topology,springel:genus}},
on the CfA2 redshift survey {\cite{vogeley:topology}} and, in a
modified way, on the nearly two--dimensional LCRS {\cite{colley:two}}.
With galaxy clusters, scales of several hundred \hMpc\ have been
probed using different morphological statistics
{\cite{coles:cluster,plionis:clusteringII}}, including the Minkowski
functionals {\cite{kerscher:abell}}.

Among this plethora of methods, Minkowski functionals currently
represent the most comprehensive description of the topological and
geometric properties of a three--dimensional structure.  Minkowski
functionals include both the percolation analysis and the genus curve.
Minkowski functionals can be applied both to density fields
{\cite{schmalzing:beyond}} and to discrete point sets
{\cite{mecke:robust}}, where the latter approach has the advantage of
introducing just one single diagnostic parameter (see below for
details), rather than, for example, a smoothing length and a threshold
as in the genus statistics.  Because Minkowski functionals are
additive they are robust measures even for sparse samples.  Minkowski
functionals also discriminate different cosmological models
efficiently {\cite{schmalzing:disentanglingI}}.  Finally, Minkowski
functionals can be used to define a planarity $\cP$ and a
filamentarity $\cF$ and thus quantify our intuition about the shape of
a structure {\cite{sahni:shapefinders}}.

Here, we study the {\em redshift space} topology of the nearby
Universe by analysing the Updated Zwicky Catalogue (UZC,
{\pcite{falco:uzc}}) which contains the redshifts of almost 20,000
galaxies and is 96\% complete in more than one fifth of the sky to the
limiting Zwicky magnitude $m_{\text{Zw}}=15.5$.  The UZC includes the
CfA2 Redshift Survey
{\cite{huchra:cfa2south,huchra:cfa2s2,huchra:cfa2s1,geller:mapping,huchra:cfa1,lapparent:slice,davis:surveyII}}.
In our analysis, we use the Minkowski functionals and their associated
shape quantifiers planarity $\cP$ and filamentarity $\cF$ which we
define in Appendix \ref{sec:mixture}.  Recently, a similar analysis on
the nearby Universe has been performed by using the LCRS
{\cite{bharadwaj:evidence}}.  The main advantage of using the UZC over
the LCRS is the geometry of the survey: the LCRS consists of six
separated slices only 1\fdg5 thick; thus, an analysis of the full
three--dimensional structure cannot be performed.  In fact,
{\scite{bharadwaj:evidence}} are unable to discriminate whether the
``filaments'' seen in the LCRS are actually sections of ``sheets''.
Because the UZC consists of two coherent regions covering $\sim14$\%
and $\sim7$\% of the sky respectively without discontinuities, we can
address this point here. In fact, {\scite{geller:mapping}} were the
first to address the two--dimensionality of the structures by
identifying the ``Great Wall'' in the CfA2 redshift survey.  On the
other hand, with the UZC, we can only probe the topology of a volume
70\hMpc\ deep, which is too small to be representative of the
Universe.

We compare the UZC redshift space topology with mock galaxy redshift
surveys extracted from $N$--body simulations of a Cold Dark Matter
(CDM) model including galaxies modelled with semi--analytical techniques 
{\cite{kauffmann:clusteringI}}.  Therefore the density bias, the
distribution of galaxies relative to the dark matter, is taken
self--consistently into account by including the physics of galaxy
formation and evolution explicitly.

In Section~\ref{sec:method} we review our analysis method.
Section~\ref{sec:uzc} briefly describes the properties of the UZC
catalogue and presents the results of our topological analysis.  We
compare the northern region of the CfA2 redshift survey with the mock
catalogues in Section~\ref{sec:mock}.  We summarise in
Section~\ref{sec:discussion}.  Two appendices provide important but
gory details of Minkowski functional analysis.

\section{Method}
\label{sec:method}

\subsection{Minkowski functionals}

\begin{table}
\makeatletter\ifSFB@referee\vspace*{10cm}\fi\makeatother
\begin{center}
\begin{tabular}{clcc}
\hline
 & geometric quantity & $\mu$ & $V_\mu$ \\
\hline
$V$    & volume               & 0 & $V$      \\
$A$    & surface              & 1 & $A/6$    \\
$H$    & mean curvature       & 2 & $H/3\pi$ \\
$\chi$ & Euler characteristic & 3 & $\chi$   \\
\hline
\end{tabular}
\end{center}
\caption{
\label{tab:minkowski}
Minkowski functionals $V_\mu$ expressed in terms of the corresponding
geometric quantities.}
\end{table}

\begin{figure}
\bildchen{\linewidth}{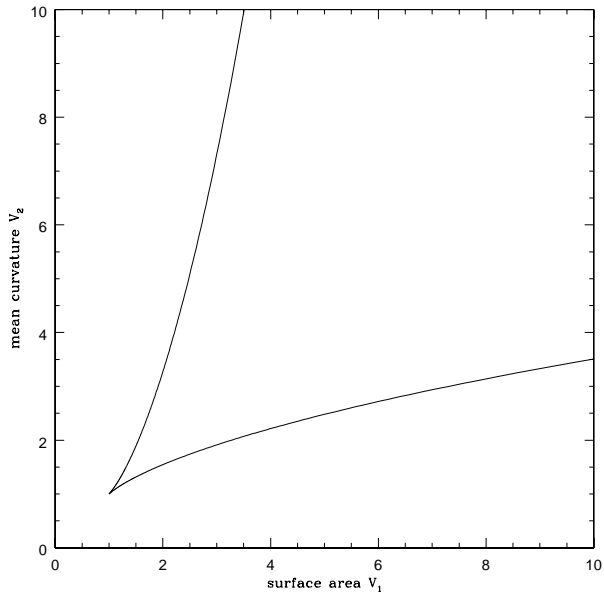}
\caption{
\label{fig:simple}
Surface area and the integrated mean curvature, namely the Minkowski
functionals $V_1$ and $V_2$, for spheroids of varying axis ratio.
Prolate spheroids (``cigars'') are characterised by large surface area
and comparatively small mean curvature, while the situation is
reversed for oblate spheroids (``pancakes'').  }
\end{figure}

Minkowski functionals are named after {\scite{minkowski:volumen}}, in
acknowledgment of his pioneering contributions to integral geometry
(for an up--to--date overview of the theory see, for example,
{\pcite{schneider:brunn}}).  {\scite{mecke:robust}} introduced the
Minkowski functionals into cosmology as descriptors of the geometry
and topology of a distribution of points.  One appealing property of
the Minkowski functionals is their uniqueness and completeness as
morphological measures, which follows by the theorem of
{\scite{hadwiger:vorlesung}} on very general requirements.
Furthermore, the four Minkowski functionals in three dimensions can be
interpreted as well--known geometric quantities, as summarised in
Table~\ref{tab:minkowski}.

Figure~\ref{fig:holes} shows examples of bodies with different Euler
characteristic $V_3$.  The Euler characteristic is a purely
topological quantity and measures the connectivity of a set.  It is
equal to the number of parts minus the number of holes.  Hence the egg
(left panel) has $V_3=1$, while the single (middle panel) and double
loop (right panel) have $V_3=0$ and $V_3=-1$, respectively.  The
behaviour of the surface area $V_1$ and mean curvature $V_2$ is
illustrated in Figure~\ref{fig:simple} on a simple family of
spheroids, whose Minkowski functionals are known analytically
{\cite{hadwiger:altes}}.  By varying the axis ratio of the spheroids
while keeping the volume fixed, we obtain a curve in the
$V_1$--$V_2$--plane.  Values are given in units of the corresponding
functionals of a sphere of the same volume, so we are indeed showing
the effects of shape.  A sphere would lie at the cusp at $(1,1)$.
Prolate spheroids (``cigars'') are characterised by large surface area
and comparatively small mean curvature, while the situation is
reversed for oblate spheroids (``pancakes'').

We consider a volume--limited subsample from a galaxy catalogue as a
set of points $\{\bx_i\}$, where $i=1\ldots{N}$.  Since the geometry
of single points is trivial, we decorate each point with a ball $B$ of
radius $r$, and measure the Minkowski functionals of the resulting
union set, the so--called Boolean grain model
{\cite{wicksell:corpuscle}}.  In short, we construct the union set 
\begin{equation}
A_r=\bigcup_{i=1}^NB_{\bx_i}
\end{equation}
and look upon the quantities $V_\mu\left(A_r\right)$ as functions of
the ball radius $r$.  By increasing the size of the balls, connections
are established between points farther apart, until the survey volume
is completely filled by the Boolean grain model.  The radius can
therefore be used as a diagnostic parameter.

\subsection{Incomplete sky coverage and finite survey depth}
\label{sec:incomplete}

\begin{figure}
\bildchen{\linewidth}{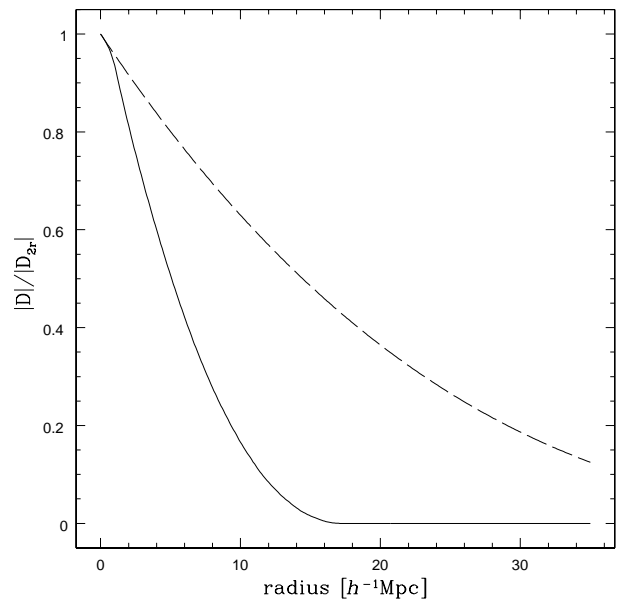}
\caption{
\label{fig:reduction}
The volume fraction of the reduced window as a function of the
shrinking distance.  The dashed line corresponds to full sky
coverage of $4\pi$; the solid line is for the CfA2n
window.  }
\end{figure}

Currently, estimation of geometrical quantities such as Minkowski
functionals relies on homogeneous sampling of the underlying point
distribution.  Galaxy catalogues, however, are usually
magnitude--limited, and therefore intrinsically inhomogeneous.  The
standard solution is to use a series of volume--limited subsamples.
However, even with volume--limited subsamples, care needs to be taken
to obtain meaningful results that are not affected by the sample
geometry.  We summarise our approach here.  An extensive description
can be found in Appendix~\ref{sec:partial}.

For calculating the volume functional $v_0$, we use the unbiased
estimator
\begin{equation}
v_0 = \frac{V_0(A_r\cap{D}_r)}{V_0(D_r)},
\end{equation}
where $D_r$ denotes the part of the volume, or {\em window}, $D$ that
is farther than $r$ from the boundary.  For all other Minkowski
functionals, we use minus estimators constructed from partial
Minkowski functionals.  An unbiased estimator of the volume densities
$v_\mu$ of the Minkowski functionals is obtained through
\begin{equation}
v_\mu=\frac{1}{|D_{2r}|}\sum_{i=1}^N\chi_{D_{2r}}(\bx_i)V_\mu(A_r;\bx_i)
\qquad(\mu=1,2,3),
\end{equation}
where $D_{2r}$ denotes the reduced sample window, and $\chi_{D_{2r}}$
is its characteristic function.

This approach works for arbitrary sky coverage, including the
non--convex boundaries of the CfA2 survey.  Moreover, at any radius,
we use all available points, since points outside the reduced sample
window still contribute as neighbours of points remaining inside. The
only drawback of our approach, when compared to the method of
{\scite{schmalzing:minkowski}}, is that the window now shrinks by
twice the radius.  This fact considerably reduces the maximum allowed
size of the Boolean ball.

We may quantify the loss of accuracy by the ratio $|D_{r}|/|D|$
between the volumes of the reduced and the original windows. For
complete sky coverage, we have
\begin{equation}
\frac{|D_{r}|}{|D|}=\left(1-\frac{r}{R}\right)^3
\end{equation}
where $R$ gives the depth of the volume--limited sample and $r$ is the
distance to be kept from the boundary.  If the sky coverage is not
complete, this fraction decreases rapidly.  This is illustrated in
Figure~\ref{fig:reduction}, which compares the reduced window size for
full sky coverage, and for the 10\% sky coverage of the CfA2n window
(defined in Section \ref{sec:models}).  Apparently, analysing the
CfA2 catalogue with Minkowski functionals becomes impractical at radii
above 10\hMpc.  We set our maximum Boolean ball radius at 5\hMpc, so
our results are not affected by this problem.

\section{The Updated Zwicky Catalogue}
\label{sec:uzc}

We now briefly describe the properties of the Updated Zwicky Catalogue
(UZC, Section~\ref{sec:description}).  In Section~\ref{sec:topology}
we compute its Minkowski functionals with the method based on the
Boolean grain model.

\subsection{Description}
\label{sec:description}

\begin{figure}
\bildchen{\linewidth}{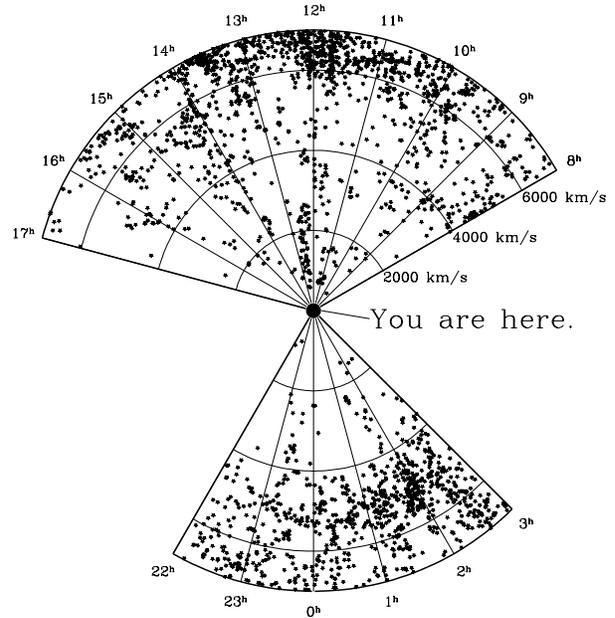}
\caption{
\label{fig:cohen.uzc}
Cone diagram of the volume--limited subsample of 70\hMpc\ depth we
analyse here from the UZC.  The larger and smaller cones correspond to
the northern and southern region, respectively.  The large dot marks
the location of our galaxy and alludes to
{\protect\scite{adams:hitchhiker}}.}
\end{figure}

{\scite{falco:uzc}} provide new measurements of the positions and
redshifts of most galaxies in the Zwicky Catalogue, which has been the
basis for the CfA1 and CfA2 redshift surveys (see \pcite{falco:uzc} and
references therein).  The UZC represents a homogeneous and well
calibrated redshift catalogue of almost 20,000 galaxies.  Furthermore,
the catalogue is now publicly available.

The UZC covers the northern celestial hemisphere which is not obscured
by the Galactic disk; the redshift catalogue is $\sim96\%$ complete to
a limiting Zwicky magnitude $m_{\text{Zw}}=15.5$.

The CfA2 redshift surveys have been compiled over many years and this
name has indicated progressively larger areas of the sky, as the
surveys were reaching completion.  The UZC contains all of the CfA2
survey analysed in the literature to April 1999.  Following
{\scite{falco:uzc}}, we refer to the area with right ascension ranges
$8\h\le\alpha_{1950}\le17\h$ and $20\h\le\alpha_{1950}\le4\h$ and
declination range $-2\fdg5\le\delta_{1950}\le50\degr$ as the CfA2
region.  The first right ascension range defines the CfA2N region
containing 12,082 galaxies with measured redshift.  The regions
between 3\h and 4\h and between 20\h and 22\h are sparsely populated
in the catalogue largely because of galactic obscuration.  Therefore
we restrict the CfA2S region to the 4,436 galaxy redshifts in the
right ascension range $22\h\le\alpha_{1950}\le3\h$.  Both areas used
in our analysis are more than 98\% complete.  Moreover the UZC
contains some galaxies fainter than the magnitude limit.  These
galaxies are mainly in multiplets which were unresolved by Zwicky.
They affect the properties of the galaxy distribution on very small
scales but should leave our analysis unchanged.  Note that the actual
number of available redshifts in the UZC is currently increasing, as
the catalogue is being updated.

Visual inspection shows well--defined structure in the CfA2 surveys,
with large voids and sheet--like structures.
Figure~\ref{fig:cohen.uzc} shows a cone diagram of the volume--limited
samples from CfA2N and CfA2S we analyse here.  The dense feature close
to the boundary of the northern part of the catalogue is the ``Great
Wall'' ({\pcite{geller:mapping}}, see also Figure~6 of
{\pcite{falco:uzc}}).  In CfA2S, the apparent galaxy concentration is
the Perseus--Pisces supercluster.

\subsection{Topology and geometry}
\label{sec:topology}

\begin{figure*}
\bildchen{\linewidth}{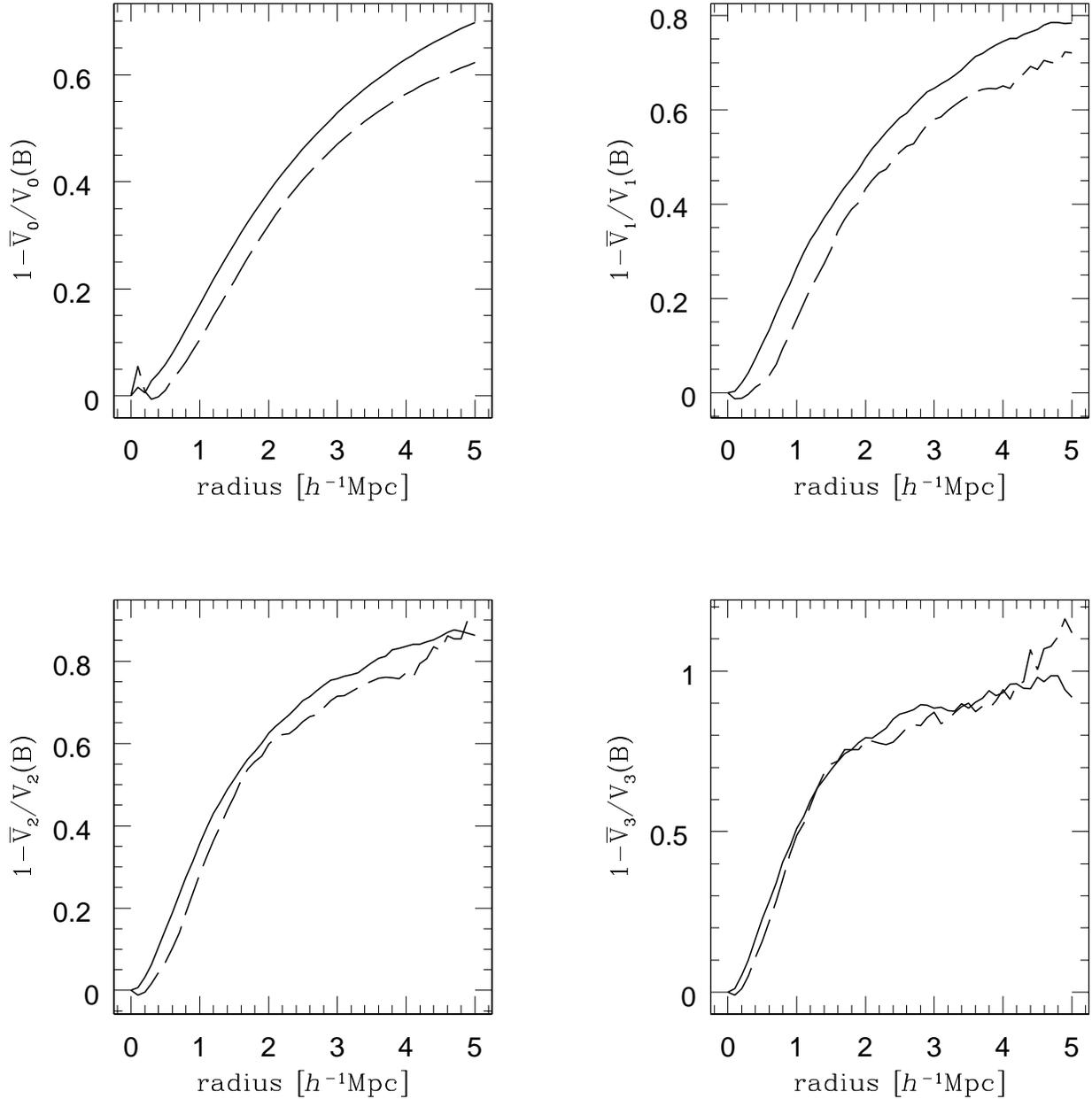}
\caption{
\label{fig:topology.depth70}
Minkowski functionals of CfA2N (solid line) and CfA2S (dashed line).
All curves were calculated from volume--limited samples 70\hMpc\
deep. Note that the CfA2N Minkowski functionals are systematically
larger than the CfA2S ones, indicating larger clustering
{\protect\cite{marzke:pairwise}}.}
\end{figure*}

In our analysis, we use volume--limited samples 70\hMpc\ deep.  This
provides a good compromise between object number (of all
volume--limited subsamples, the 70\hMpc\ one includes the highest
number of objects) and depth (parts of the Great Wall are closer than
70\hMpc).  Since the scaling behaviour of Minkowski functionals with
the number density is not known in general, we take care to always
analyse samples with the same number density of galaxies.  From all
volume--limited samples, we choose a set of slightly smaller random
samples such that the number density corresponds to exactly 1,000
galaxies in the CfA2n volume (see Section~\ref{sec:models} for a
definition).

The results of our analysis of the UZC are shown in
Figure~\ref{fig:topology.depth70}.  Instead of the Minkowski
functionals $v_\mu$ themselves, we plot the quantities
$1-\lV_\mu/V_\mu(B)$.  A thorough motivation of their use is given in
Appendix~\ref{sec:mixture}; most notably they are exactly equal to
zero for a stationary Poisson process, i.e.\ a random distribution of
points with spatially constant density.  Generally speaking, values
above zero reveal a clustered distribution of galaxies.  Note,
however, that the $\lV_\mu$ can be written as an alternating series
containing the hierarchy of correlation functions
{\cite{schmalzing:quantifying}}, so clustering does not simply mean
that the two--point function is positive.  As an example, consider
$1-\lV_0/V_0(B)$.  By Equation~(\ref{eq:poisson}) this quantity
increases monotonically with $1-v_0$, the void probability function
{\cite{white:hierarchy}}.  Hence larger values in our plots indicate
larger voids in the point distribution, which in turn means that
points have to cluster more tightly.

Some of the functionals, most notably the volume $v_0$, indicate
stronger clustering for CfA2N than for CfA2S.  In fact, CfA2N contains
more clusters than CfA2S and the Great Wall is more apparent than the
Perseus--Pisces supercluster in CfA2S.  The two--point correlation
function already showed that the CfA2N clustering is peculiarly large
when compared to other surveys
{\cite{marzke:pairwise,diaferio:clusteringIII}}.

We use a toy model to characterise the geometry of CfA2: we assume
that galaxies are distributed in one--dimensional filaments,
two--dimensional sheets or in a homogeneous field.  We can compute the
Minkowski functionals for this model analytically. Two coefficients,
$\cF$ and $\cP$, give the relative contribution of filaments and
sheets in the galaxy distribution (see Appendix~\ref{sec:mixture} for
details).  Note, however, that this approach has the main shortcoming
of assigning structures which are loosely clustered to the planarity
component, because alternative assignments would yield considerably
worse fits.  Therefore, $\cF$ and $\cP$ yield the relative importance
of the two structural components, but their exact values should be
considered cautiously.

By fitting our toy model to the CfA2N Minkowski functional curves, we
obtain $\cF=0.12$ and $\cP=0.67$ for the filamentarity and planarity
contribution, respectively.  The high degree of planarity is expected
from the visual appearance of the Great Wall in the catalogue.  For
the CfA2S we find $\cF=0.01$ and $\cP=0.75$. The large value of $\cP$
in both regions reflects the tendency of galaxies to be distributed in
two--dimensional structures or in loose structures rather than in
filaments.  Moreover, the smaller size of the CfA2S region also tends
to increase slightly the signal for planar and loose structures.

Because of its simplicity, our toy model provides a straightforward
way of measuring planarity and filamentarity, but it is inadequate
to model more complex realistic situations. We are planning some
refinements in order to obtain results in better agreement with our
intuitive interpretations of the galaxy distribution.

\section{Mock galaxy redshift surveys}
\label{sec:mock}

\begin{figure}
\bildchen{\linewidth}{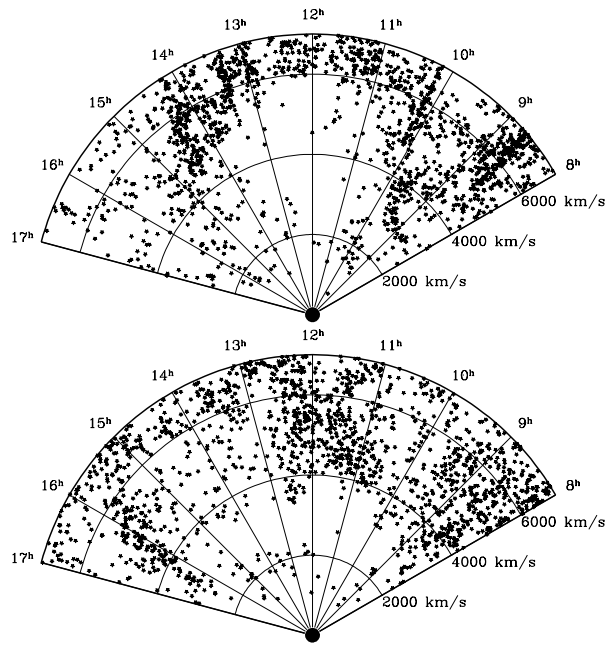}
\caption{
\label{fig:cohen.mock}
Cone diagrams of two mock catalogues.  The upper and lower cones are a
volume--limited subsample constructed from the CfALF $\Lambda$CDM
model and the CfALF $\tau$CDM model, respectively.  The observer
galaxy is indicated by the large dot. }
\end{figure}

We compare the morphology of the CfA2n redshift survey with two
variants of a CDM model.  In Section~\ref{sec:models} we briefly
describe the simulations and how we model galaxy formation and
evolution.  Section~\ref{sec:segregation} discusses how the luminosity
limit of the galaxy sample, the shape of the sample volume and the
peculiar velocity distortions in redshift space affect the Minkowski
functionals.  Section~\ref{sec:comparison} contains the comparison of
CfA2n with the mock galaxy redshift surveys.

\subsection{Models}
\label{sec:models}

We consider two simulations from the GIF project
{\cite{kauffmann:clusteringI}}: a CDM model with either $\Omega_0=1$
($\tau$CDM) or $\Omega_0=0.3$ and $\Lambda=0.7$ ($\Lambda$CDM).  The
simulations contain $256^3$ dark matter particles each.  The
normalisation $\sigma_8$ of the power spectrum of the initial density
perturbations was chosen to reproduce the present day abundance of
rich galaxy clusters.  Table~\ref{tab:gif} summarises the parameters
of the models.

\begin{table}
\makeatletter\ifSFB@referee\vspace*{10cm}\fi\makeatother
\caption{GIF Simulations}\label{tab:gif}
\begin{tabular}{rrrrrrr}
model & $\Omega_0$ & $\Omega_\Lambda$ & $h$ & $\sigma_8$ & $m_p$ & $L$ \\
$\Lambda$CDM & 0.3 & 0.7 & 0.7 & 0.90 & 1.4 & 141 \\
$\tau$CDM & 1.0 & 0.0 & 0.5 & 0.60 & 1.0 & 85 \\
\end{tabular}

\medskip
{Parameters of the two GIF simulations {\cite{kauffmann:clusteringI}}
used in this paper.  The Hubble constant $h$, the particle mass $m_p$,
and the comoving size $L$ of the simulation box are in units of
$H_0=100\kms\Mpc^{-1}$, $10^{10}\hMsun$, and \hMpc, respectively.}
\end{table}  

The simulations were run with {\em Hydra} {\cite{pearce:hydra}}, the
parallel version of the AP$^3$M code
{\cite{couchman:mesh,couchman:hydra}}, kindly provided by
{\scite{jenkins:virgo}}.

{\scite{kauffmann:clusteringI}} combine these high resolution $N$--body
simulations with semi--analytic modelling to form and evolve galaxies
within dark matter haloes: gas cooling, star formation, supernova
feedback, stellar evolution and galaxy--galaxy merging are the
relevant processes included.  At any cosmological epoch, the
observable properties of the galaxies, namely luminosity, colour, star
formation rate, and stellar mass, are the result of the galaxy's
merging history, its interaction with the environment and the passive
evolution of its stellar content.

The models predict photometric and clustering properties of galaxies
at redshift $z=0$
{\cite{kauffmann:clusteringI,diaferio:clusteringIII}}, and their
evolution to high redshift
{\cite{kauffmann:clusteringII,diaferio:clusteringIV}} which are in
reasonable agreement with observations.  Moreover, these analyses
suggest further diagnostics for constraining the cosmological model
and determining the relevance of the different galaxy formation
processes.

These simulations represent the only models, available to date, where
galaxies form and evolve in a self--consistent cosmological model and
within a volume of the Universe sufficiently large for an analysis of
the large scale galaxy distribution (see also
{\pcite{pearce:simulation}}).  In fact,
{\scite{diaferio:clusteringIII}} extract mock galaxy redshift surveys
from the simulations at $z=0$ for a direct comparison with the CfA2N
redshift survey, whose volume $\sim7\times10^5h^{-3}$Mpc$^3$ is
comparable to the volume of the simulation boxes.  Specifically, they
consider the area $8\h\le\alpha_{1950}\le17\h$,
$8\fdg5\le\delta_{1950}\le44\fdg5$ which has a declination range
smaller than CfA2N.  We refer to this region as the CfA2n region.  In
order to resemble the CfA2n appearance in the mock catalogues as
closely as possible, the mock redshift surveys were compiled by
locating the observer home galaxy within the simulation box on a
galaxy similar to the Milky Way at $\sim70\hMpc$ away from a
Coma--like massive cluster.

{\scite{diaferio:clusteringIII}} analyse both the CfA2n and the mock
redshift surveys with the same standard observational techniques. They
show that the small scale clustering properties of galaxies, namely
galaxy group properties, redshift space correlation function and
pairwise velocities, are reproduced by both models independently of
$\Omega_0$.  Despite this success, a visual inspection of the mock
redshift surveys shows remarkable differences in the large scale
distribution of galaxies.  Both models fail to produce structures as
coherent and as sharply defined as in the CfA2n survey, although
$\Lambda$CDM catalogues seem to be in better qualitative agreement
with the data (see their Figures 5--8).  
Two examples of volume--limited subsamples constructed from these
models are shown in Figure~\ref{fig:cohen.mock}.  

The $B$--band galaxy luminosity function predicted by the
semi--analytical models is not a good fit to the CfA survey luminosity
function {\cite{marzke:luminosity}}. To investigate the effect of the
luminosity function on the small--scale clustering properties,
{\scite{diaferio:clusteringIII}} assign new luminosities to the
galaxies in order to reproduce the CfA luminosity function exactly,
while preserving the galaxies' luminosity rank. We thus have two sets
of mock catalogues, where the galaxies have luminosities according to
the semi--analytical model luminosity function (SALF) or the CfA
luminosity function (CfALF).  We have analysed both the SALF and CfALF
mock catalogues.  Similarly to other measures related to the
underlying mass distribution, the Minkowski functionals are basically
independent of the luminosity function adopted.  Throughout this
analysis, we show results from the CfALF catalogues alone.  Moreover,
we consider ten mock catalogues for each simulation to assess the
robustness of our results.

\subsection{Luminosity, geometry and redshift space effect}
\label{sec:segregation}

\begin{figure*}
\bildchen{\linewidth}{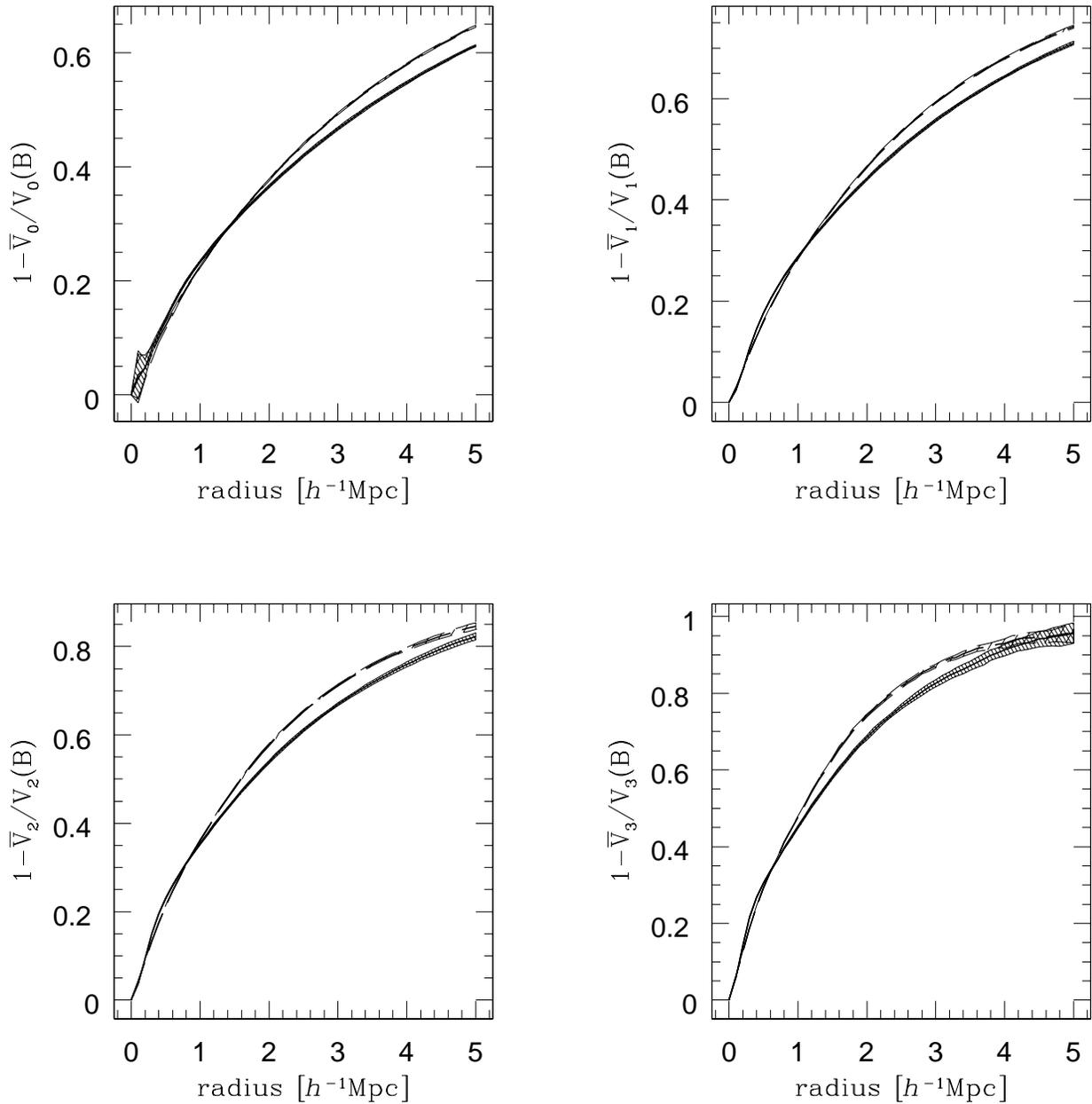}
\caption{
\label{fig:selection}
Effect of luminosity and geometry on the Minkowski functionals of the
CfALF $\Lambda$CDM model.  The dashed and solid lines are for all
galaxies in the simulation box brighter than $M_{\text{B}}=-18.7 +
5\log h$ and $M_{\text{B}}=-17.5 + 5\log h$ respectively.  The shaded
area indicates the $1\sigma$--spread over ten subsamples.  }
\end{figure*}

{\scite{kauffmann:clusteringI}} show that the two--point correlation
function is similar for galaxy samples with different magnitude
limits.  The Minkowski functionals are able to extract more
information from the galaxy distribution.  In fact, bright galaxies
cluster stronger than faint ones.  Figure~\ref{fig:selection} shows
the Minkowski functionals for the $\Lambda$CDM simulation; $\tau$CDM
yields very similar plots.  In order to keep constant the statistical
significance of different galaxy samples, we compute the Minkowski
functionals only of subsamples containing the same number density of
galaxies, and average over ten subsamples.  The solid lines show the
Minkowski functionals for galaxies brighter than $M_B=-17.5+5\log{h}$.
The Minkowski functionals for galaxies brighter than
$M_B=-18.7+5\log{h}$ are indicated by dashed lines.  This absolute
magnitude corresponds to an apparent magnitude $m_B=15.5$ at 70\hMpc.
The difference between the two curves is significantly larger than the
scatter between different subsamples from the same set of galaxies.

\begin{figure*}
\bildchen{\linewidth}{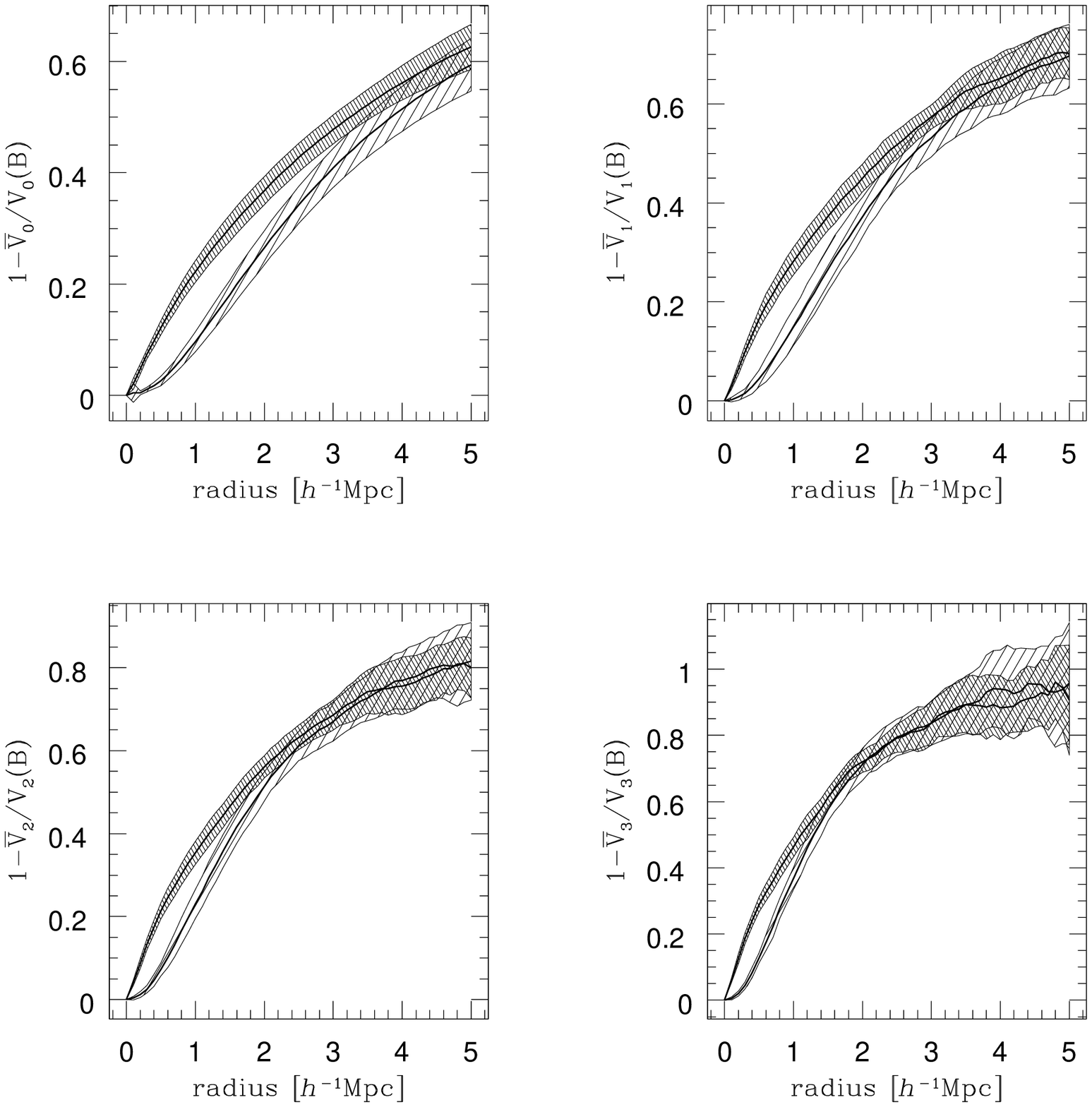}
\caption{
\label{fig:realred.depth70}
Minkowski functionals of the $\Lambda$CDM mock catalogues in real
space (dark shaded area) and in redshift space (light shaded
area). The shaded area indicates the r.m.s.\ spread over the ten mock
catalogues. }
\end{figure*}

We have also checked that the Minkowski functionals computed with our
procedure (Section~\ref{sec:incomplete}) are insensitive to the shape
of the sample volume.  We do this by calculating the Minkowski
functionals of the real space position of the galaxies averaged over
the volume--limited subsamples extracted from the ten mock redshift
surveys.  The dashed lines in Figure~\ref{fig:selection} lie within
the scatter of the resulting Minkowski functionals (shown as the dark
shaded area in Figure~\ref{fig:realred.depth70}).  Since they come
from volumes which have differing shape (periodic, rectangular box
versus CfA2n volume ) but contain galaxies with roughly the same
luminosity, we can deduce that the difference due to the sample
geometry is negligible in comparison to the scatter between individual
mock catalogues.  Note that for mock catalogues, the scatter is much
larger than for subsamples from the whole simulation box, simply
because a mock catalogue contains fewer galaxies.

Redshift space distortions tend to destroy structure on small scales,
because groups and clusters in real space are diluted in redshift
space by the finger--of--god effect.  Figure~\ref{fig:realred.depth70}
shows this difference in the $\Lambda$CDM model.  The Minkowski
functionals of the real and redshift space galaxy distributions almost
coincide at larger values of the Boolean ball radius, because redshift
space distortions become less relevant on large scale.

\subsection{Comparison with the CfA2n}
\label{sec:comparison}

\begin{figure*}
\bildchen{\linewidth}{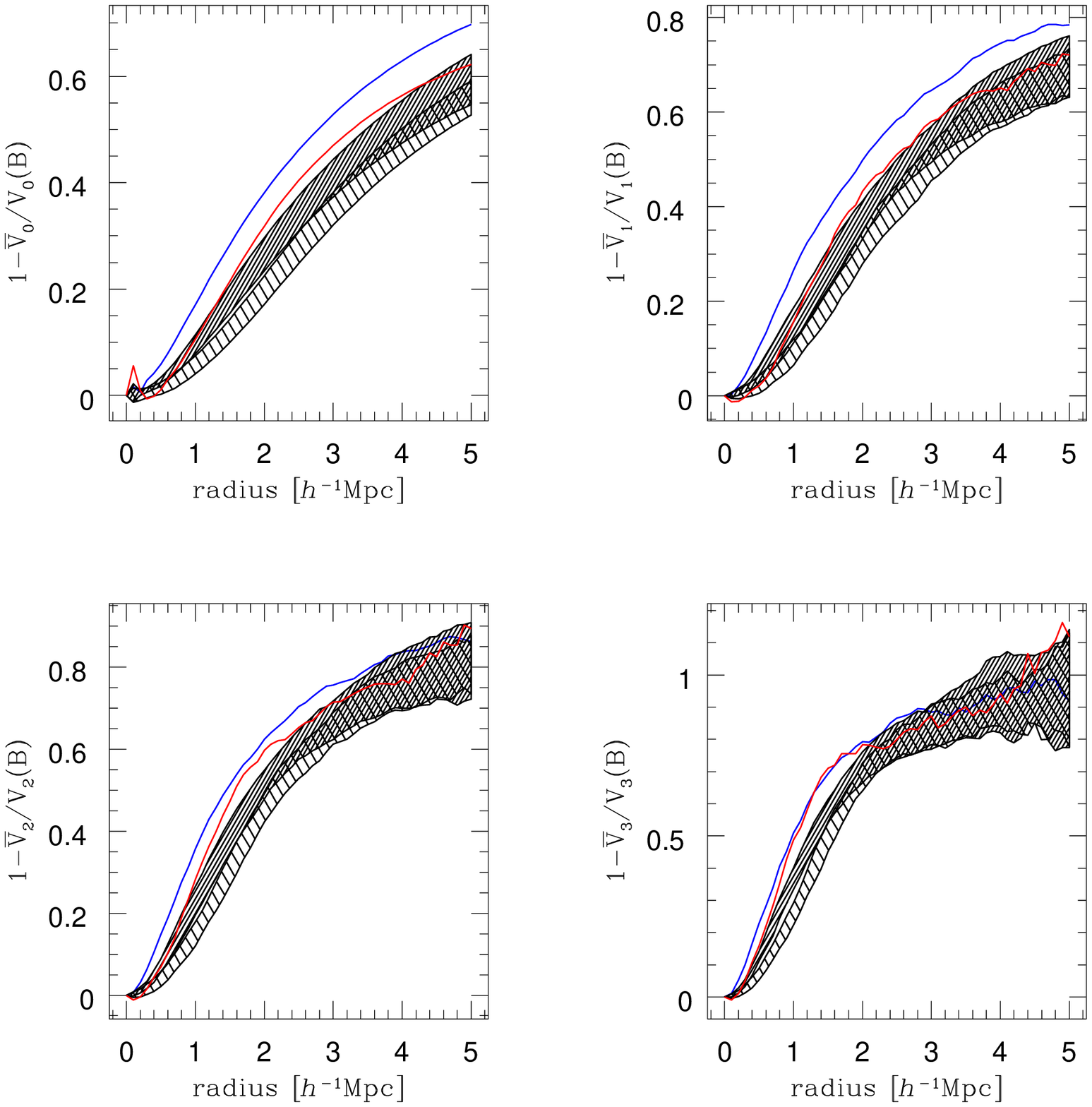}
\caption{
\label{fig:structure.depth70}
Structure in the CfA2n redshift survey, and in the mock catalogues
extracted from the $\Lambda$CDM (dark shaded area) and the $\tau$CDM
(light shaded area) models.  The areas indicate the 1--$\sigma$ scatter
over the ten mock catalogues constructed for each model.  All curves
are for volume--limited samples 70\hMpc\ deep.}
\end{figure*}

Figure~\ref{fig:structure.depth70} compares the morphological
properties of CfA2n with the mock catalogues: neither of the two
models is capable of reproducing the CfA2n properties.  At Boolean
ball radii $\sim1-2\hMpc$, all the four Minkowski functionals of CfA2n
are five times the r.m.s.\ above the average mock catalogue of the
$\Lambda$CDM.  Given the rather small number of ten realisations, we
can conclude that the CfA2n clustering is not reproduced by the
$\Lambda$CDM model at the 90\% confidence level.  The $\tau$CDM model
shows an even smaller degree of clustering.  At larger radii and for
the mean curvature $v_2$ and Euler characteristic $v_3$, the
disagreement between the models and observations is less dramatic, but
also less significant, because the sample volume is filled by a
considerably smaller number of Boolean balls.  It is crucial to
recall, however, that the scatter of quantities derived from the
variance over all ten mock catalogues underestimates the sampling
variance, because the mock surveys are constructed from the same small
parent simulation.

The failure of the $\Lambda$CDM in reproducing the CfA2n Min\-kows\-ki
functionals is particularly interesting, because, on average, this
model fits other observations reasonably well and in any case better
than the $\tau$CDM model
{\cite{kauffmann:clusteringI,kauffmann:clusteringII}}.  The Minkowski
functionals confirm that $\Lambda$CDM should be preferred to
$\tau$CDM.

It is worth noting that, because of the smaller degree of clustering
of CfA2S when compared to CfA2N (see
Figure~\ref{fig:topology.depth70}), the difference between the CfA2S
and the models is less pronounced, although still significant.  The
result is an example of the power of the Minkowski functionals
analysis. In fact, \scite{diaferio:clusteringIII} show that the two
point correlation function of the mock catalogues agrees quite well
with the CfA2S function. As expected, the Minkowski functionals are
able to extract more information on the clustering properties of
galaxies.
 

We can also perform the fitting procedure described in
Appendix~\ref{sec:mixture} to estimate the planarity and filamentarity
of the structure contained in the mock catalogues.  $\Lambda$CDM
yields $\cF=0.01$ and $\cP=0.64$, while $\tau$CDM results in
$\cF=0.02$ and $\cP=0.60$ for the average mock catalogues.  As pointed
out in Section \ref{sec:topology}, the large values of $\cP$ compared
to $\cF$ indicate that the galaxies tend to be distributed in
two--dimensional or loose structures rather than filaments.  Compared
to the values determined from the real CfA2N catalogue in
Section~\ref{sec:topology} ($\cF=0.12$ and $\cP=0.67$), $\cP$ is $\sim
10\%$ smaller in the models than in CfA2N, again indicating a smaller
degree of clustering.

We cannot exclude that the failure of the models in reproducing the
Minkowski functionals of CfA2N is due to the small size of both the
simulation box and the real survey.  In fact, the initial power
spectrum of the density perturbations does not contain Fourier
components with wavelength larger than the box size.  This prevents
the development of large--scale non--Gaussian features such as the
Great Wall.  Moreover, as already indicated by an analysis of the IRAS
1.2Jy catalogue {\cite{kerscher:fluctuations}}, a volume--limited
subsample 70\hMpc\ deep is too small to be a fair sample of the
Universe.  Therefore, our analysis is likely to be dominated by local
effects, which should average out when deeper catalogues are
considered.  This hypothesis could already be tested either with
simulations of larger volumes (see
e.g. {\pcite{doroshkevich:superlargescale}}), where one could estimate
the cosmic variance, or with a constrained realisation
{\cite{mathis:}}, which should not suffer from the fair sample
problem.

\section{Conclusion}
\label{sec:discussion}

{\scite{lapparent:slice}} and {\scite{geller:mapping}} first pointed
out that the large--scale galaxy distribution in the CfA2 redshift
survey has a two--dimensional sheet--like structure rather than a
one--dimensional filamentary appearance.  Here we quantify this visual
impression by analysing the redshift space topology and geometry of
the UZC.  By constructing a Boolean grain model of balls centred
around the galaxies, we obtain unbiased estimates for the Minkowski
functionals of the galaxy distribution.  We then fit the Minkowski
functionals to a toy model to extract quantitative information on the
degree of filamentarity and planarity in CfA2.  We find that
filament-- and sheet--like structures contribute 12\% and 67\% to the
Minkowski functionals of CfA2N respectively, while the remainder comes
from a homogeneous distribution of galaxies.  Our results therefore
strongly suggest that the filaments identified in the LCRS
{\cite{bharadwaj:evidence}} are actually sections of sheets.

Apparently, our simple model is inadequate in other realistic
situations, such as CfA2S (the southern part of the UZC), where looser
structures are present.  However, the encouraging result we obtain
with CfA2N indicates that it is worth pursuing this approach with some
refinements.  When deeper surveys as the Sloan Digital Sky Survey will
be available, our approach can represent a valuable tool to quantify
the topology of the galaxy distribution.  Most notably, contrary to
{\scite{sahni:shapefinders}} or {\scite{schmalzing:disentanglingI}},
the method presented here does not rely on shape measurements of
isolated isodensity contours.  Hence we can also obtain meaningful
results in low to intermediate density environments, where objects in
a smoothed field already percolate and are inaccessible to
measurements.

We also compare CfA2n, a smaller volume of CfA2N, with mock redshift
surveys extracted from a $\Lambda$CDM and a $\tau$CDM model.  The
models do not show the high degree of clustering of CfA2n. However,
CfA2n seems to be quite a peculiar region of the Universe and the
disagreement between the models and, for example, CfA2S is less
dramatic. Moreover, the Minkowski functionals of the $\Lambda$CDM are
closer to observations than the Minkowski functionals of $\tau$CDM.
In conclusion, although our topological analysis confirms earlier
results which indicate that overall $\Lambda$CDM is a better
representation of the real Universe than $\tau$CDM, we show, at the
same time, that $\Lambda$CDM is not yet a satisfactory model of the
Universe.
  
{\scite{vogeley:topology}} compare the genus statistics of CfA2 with a
$\Lambda$CDM model. They find that this model is consistent with CfA2
on scales $\ge10\hMpc$, but it fails to reproduce CfA2 on smaller
scales.  They use a simple bias scheme to identify galaxies and
extract mock catalogues from $N$--body simulations. So the physics
acting on small scales, which is missing from their simulations, might
be responsible for the disagreement. Here, we used semi--analytic
modelling to form and evolve galaxies in $N$--body simulations, but
still the agreement between $\Lambda$CDM and CfA2 does not improve.

Note that our Boolean grain model approach follows a direction
complementary to the genus statistics analysis generally used.  The
standard method introduced by {\scite{gott:sponge}} employs a density
field smoothed on varying scales.  Increasing the smoothing kernel
width successively erases structure on small scales
{\cite{koenderink:scalespace}}.  The analysis of
{\scite{vogeley:topology}}, for example, employs widths between
6\hMpc\ and 20\hMpc\ and addresses the issue of the Gaussianity of the
distribution of the initial density perturbations.  We adopt the
Boolean grain model approach by {\scite{mecke:robust}} and decorate
each galaxy with a sphere of radius $r$ varying from zero to $5\hMpc$.
This procedure captures the morphology of distinct structures on very
large scales while preserving the small scale structure information.

Finally, the small volume of both the survey and our simulation boxes
represents a severe limit to our analysis, because we are unable to
quantify the effect of cosmic variance.  In a forthcoming paper, we
will use simulation boxes of 320\hMpc\ and 480\hMpc\ on a side to
quantify these effects.  We will also extract mock redshift surveys of
the Sloan Digital Sky Survey {\cite{gunn:sdss}} and the 2dF survey
{\cite{maddox:2df}}.  We will use a simple biasing scheme to locate
galaxies (e.g. {\pcite{cole:mock}}), rather than the semi--analytic
modelling we have used here.  With these larger volumes, we should be
able to use our topological approach for discriminating between
cosmological models successfully.

\section*{Acknowledgments}

We thank Emilio Falco and his collaborators for their effort in
homogenising the Zwicky catalogue and making the UZC data available.
We thank Margaret Geller and Simon White for relevant suggestions and
Claus Beisbart, Thomas Buchert, Martin Kerscher, and Volker Springel
for interesting discussions and valuable comments.  The $N$--body
simulations were carried out at the Computer Center of the Max--Planck
Society in Garching and at the EPPC in Edinburgh, as part of the Virgo
Consortium Project.  AD acknowledges support from an MPA guest
post--doctoral fellowship.

\appendix

\section{Partial Minkowski functionals}
\label{sec:partial}

Throughout our article, we consider the Minkowski functionals of a
Boolean grain model $A_r$ constructed by placing balls of radius $r$
around a set $\{\bx_i\}$ of $N$ points contained in a spatial region
$D$;
\begin{equation}
A_r=\bigcup_{i=1}^NB_r(\bx_i).
\end{equation}
For comparing samples of different size, we may define volume
densities $v_\mu$ of the Minkowski functionals of this objects by
\begin{equation}
v_\mu=\frac{1}{|D|}V_\mu(A_r),
\label{eq:density}
\end{equation}
if $D$ is a large box with periodic boundary conditions, and $|D|$
denotes its volume.

Unfortunately, for many practical applications, the sample volume is
not a box and does not feature periodic boundaries.  Therefore, one
must use unbiased estimators for calculating the Minkowski
functionals; such a method of removing the boundary effects given
arbitrary sample geometries is provided through partial Minkowski
functionals.

First of all, to take into account the boundary effects on the volume
functional $v_0$, the sample window $D$ must be reduced by the radius
$r$ to obtain an unbiased estimate.  If $D_r$ denotes the set of all
points that are farther than $r$ from the boundary of the sample
window $D$, we have
\begin{equation}
v_0 = \frac{V_0(A_r\cap{D}_r)}{V_0(D_r)}.
\end{equation}
{\scite{maurogordato:void}} introduced this estimator for calculating
the void probability function of the CfA1 catalogue.  An extension was
made by {\scite{schmalzing:minkowski}} to include all Minkowski
functionals of a point set in a convex window.

While for a smooth body, all Minkowski functionals $v_\mu$,
$\mu=1,2,3$ can be written as surface integrals, this is no longer
possible for the Boolean grain model, which has edges and corners at
intersections of two or more balls.  Nevertheless the contributions of
these non--regular surface to the Minkowski functionals may still be
evaluated using a more general concept of surface integration that
includes edges and corners {\cite{federer:curvature,schneider:brunn}}.
Following {\scite{mecke:robust}}, we arrive at the so--called
partition formula, which sorts contributions to the Minkowski
functionals from the various types of boundary, i.e. surface points
belonging to one, two or three balls\footnote{Although intersections
of four and more points may occur for very special point distributions
{\protect\cite{platzoeder:diplom}}, they are irrelevant since they
provide no contribution on average.  In practice, slight displacements
of the order of the numerical resolution can be used to resolve such
situations without losing accuracy.}.  The partition formula states
that
\begin{multline}
V_\mu(A_r)
= \\
\sum_{i=1}^N V_\mu^{(i)}(A_r)
+\frac{1}{2}\sum_{i,j=1}^N V_\mu^{(ij)}(A_r)
+\frac{1}{6}\sum_{i,j,k=1}^N V_\mu^{(ijk)}(A_r).
\end{multline}
$V_\mu^{(i)}(A_r)$ denotes the contribution from the uncovered surface
of the ball around point $\bx_i$, while $V_\mu^{(ij)}(A_r)$ comes from
the intersection line of balls $i$ and $j$, and $V_\mu^{(ijk)}(A_r)$
is located at the corner made by the intersection of balls $i$, $j$
and $k$.  Rearranging the summations, we have
\begin{equation}
V_\mu(A_r)
=
\sum_{i=1}^N V_\mu(A_r;\bx_i),
\label{eq:partial}
\end{equation}
where the partial Minkowski functional $V_\mu(A_r;\bx_i)$ sums up
contributions that include the point $i$ located at $\bx_i$.  Since an
intersection, and hence a non--zero contribution, is only possible if
all balls are less than $2r$ apart, the partial Minkowski functionals
measure the local morphology in a well--defined neighbourhood, namely a
ball of radius $2r$, around each point.

Equation~(\ref{eq:partial}) is already useful for the practical
evaluation of Minkowski functionals for a set of points in a
rectangular box with periodic boundary conditions.  For detailed
information, the reader is referred to the source code of the program
that is available from the authors.  Basically, one calculates all
partial Minkowski functionals and sums them up to estimate the density
through Equation~(\ref{eq:density}), by
\begin{equation}
v_\mu=\frac{1}{|D|}\sum_{i=1}^N V_\mu(A_r;\bx_i).
\end{equation}

Even for a complicated survey geometry, the global Min\-kows\-ki
functionals of the Boolean grain model can now be written as sums of
the partial Minkowski functionals.  Since only neighbours within $2r$
around a point contribute to its partial Minkowski functionals, we can
simply restrict the summation to the part of the sample that is
further than $2r$ from the boundary.  Calling this shrunken window
$D_{2r}$, we have
\begin{equation}
v_\mu=\frac{1}{|D_{2r}|}\sum_{i=1}^N\chi_{D_{2r}}(\bx_i)V_\mu(A_r;\bx_i),
\end{equation}
where 
\begin{equation}
\chi_{D_{2r}}(\bx)=\left\{\begin{array}{c}
1\text{ if }\bx\in D_{2r} \\
0\text{ if }\bx\not\in D_{2r}
\end{array}
\right.
\end{equation}
is its characteristic function.  In the language of spatial
statistics, this quantity is a minus estimator for the volume
densities of the Minkowski functionals.  Minus estimators especially
for the two--point correlation function are already known in
cosmology; they have been used for example by
{\scite{coleman:fractal}}, and thoroughly investigated by
{\scite{kerscher:twopoint}}.

One should keep in mind that minus estimators provide unbiased
estimates only if applied to a sample from a {\em stationary} point
process.  Hence we always use volume--limited subsamples from
catalogues when carrying out our analysis.

\section{Definition of filamentarity and planarity through a toy model}
\label{sec:mixture}

\begin{figure}
\bildchen{\linewidth}{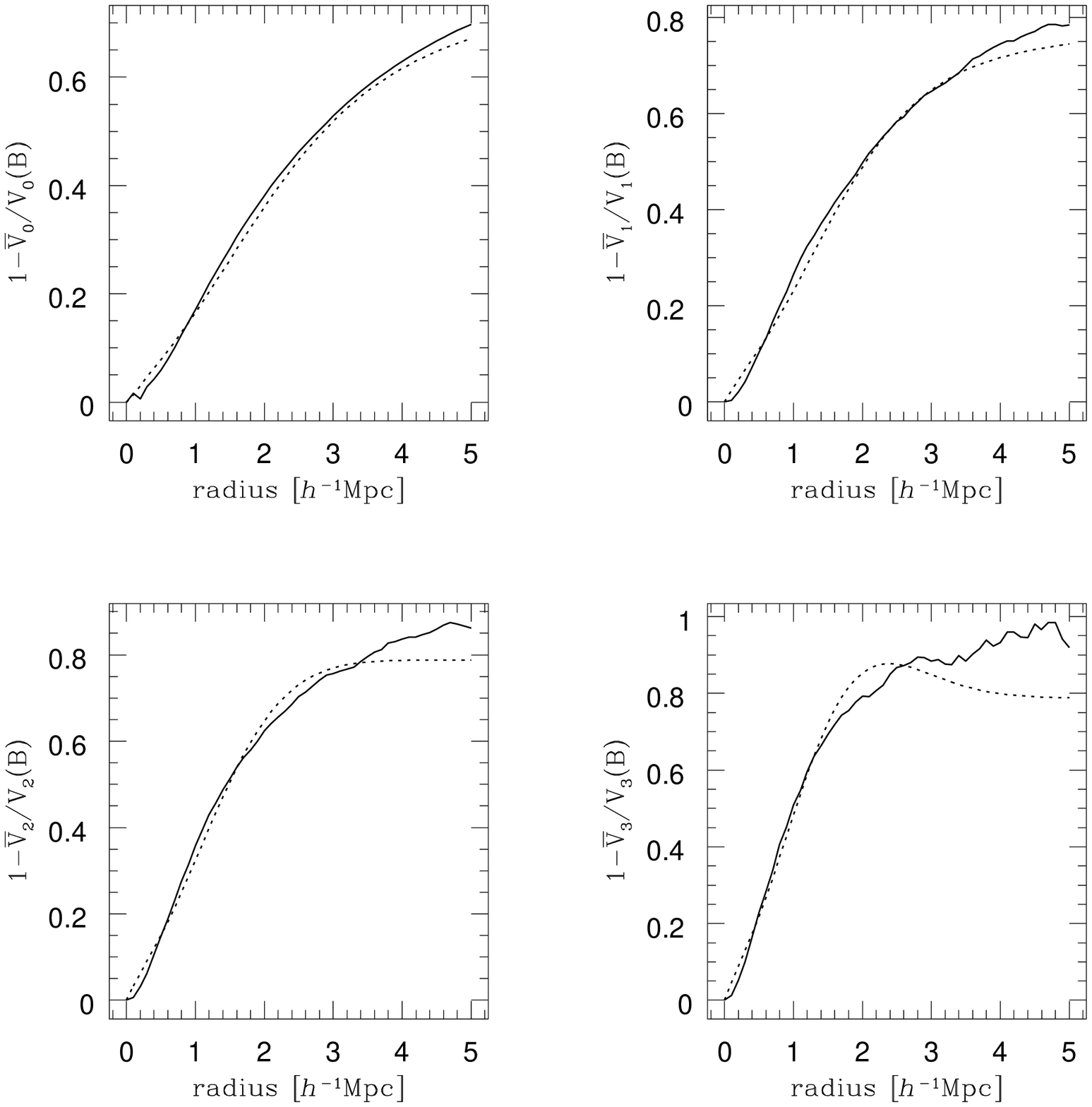}
\caption{
\label{fig:fit}
The Minkowski functionals of CfA2N (solid line) compared with the best
fit of our toy model (dotted line).
}
\end{figure}

Consider a random mixture of $N$ bodies with {\em different} Minkowski
functionals.  The average Minkowski functionals of the resulting union
are given by {\cite{mecke:diss}}
\begin{equation}
\begin{split}
v_0 &= 1-\re^{-\rho\lV_0} \\
v_1 &= \re^{-\rho\lV_0} \rho\lV_1 \\
v_2 &= \re^{-\rho\lV_0} \left( \rho\lV_2-\frac{3\pi}{8}\rho^2\lV_1^2 \right) \\
v_3 &= \re^{-\rho\lV_0} \left( \rho\lV_3-\frac{9}{2}\rho^2\lV_1\lV_2+\frac{9\pi}{16}\rho^3\lV_1^3 \right)
\end{split}
\label{eq:poisson}
\end{equation}
where $\lV_\mu$ is the number weighted average over the isolated
Minkowski functionals of all different bodies.

Although the Boolean grain model used in our analysis places identical
balls around all points, we may use the formula above for calculating
analytically the Minkowski functionals of a simple toy model.  In
fact, we can take into account the inhomogeneous distribution of
identical balls by considering them as randomly distributed objects
with Minkowski functionals characteristic of their respective
environment.

Let us assume that points are located in one--dimensional filaments,
in two--dimensional sheets, or in a homogeneous field.  If each of
these types of structure were isolated from the rest, we could
calculate the Minkowski functionals per single point analytically
{\cite{schmalzing:diplom}}.  For the average point in the filament,
sheet and field, respectively, we have\footnote{We have that
$\Psi(x)=\frac{2}{\sqrt{\pi}}\int_0^{x}\dt\exp(t^2)$ and $I_0(x)$ is
the modified Bessel function of order zero.  }
\begin{equation}
\begin{split}
\frac{\lV_0^{(\text{fil})}}{V_0(B)} &= 
\tfrac{3}{4}\frac{x+\re^{-2x}}{x^2}-\tfrac{3}{8}\frac{1-\re^{-2x}}{x^3}
,\\
\frac{\lV_1^{(\text{fil})}}{V_1(B)} &= 
\tfrac{1}{2}\frac{1-\re^{-2x}}{x}
,\\
\frac{\lV_2^{(\text{fil})}}{V_2(B)} &= 
\tfrac{1}{2}\frac{1-\re^{-2x}}{x}-x\int_0^1\!\dt\re^{-2xt}\sqrt{1-t^2}\arcsin{t}
,\\
\frac{\lV_3^{(\text{fil})}}{V_3(B)} &=
e^{-2x}
,
\end{split}
\end{equation}
\begin{equation}
\begin{split}
\frac{\lV_0^{(\text{sheet})}}{V_0(B)} &= 
\tfrac{3}{2\pi}\frac{1}{x^2}-\tfrac{3}{4\pi}\frac{\re^{-{\pi}x^2}\Psi(\sqrt{\pi}x)}{x^3},\\
\frac{\lV_1^{(\text{sheet})}}{V_1(B)} &= 
\tfrac{1}{2}\frac{\re^{-{\pi}x^2}\Psi(\sqrt{\pi}x)}{x}
,\\
\frac{\lV_2^{(\text{sheet})}}{V_2(B)} &= 
\tfrac{1}{2}\frac{\re^{-{\pi}x^2}\Psi(\sqrt{\pi}x)}{x}-\tfrac{\pi}{2}x^2\re^{-{\pi}x^2}\\&\quad\times\int_0^1\!\dt\,\exp(\frac{\pi}{2}x^2t^2)\,I_0(\frac{\pi}{2}x^2t^2)\,t^2\,\arcsin{t}
,\\
\frac{\lV_3^{(\text{sheet})}}{V_3(B)} &=
e^{-{\pi}x^2}(1-{\pi}x^2)
,
\end{split}
\end{equation}
\begin{equation}
\frac{\lV_\mu^{(\text{field})}}{V_\mu(B)} = 1, \qquad(\mu=0,\dots,3) 
\end{equation}
where we have normalised the functionals by dividing by the values for
an isolated ball of the same radius, and $x=r/d$ denotes the radius
divided by the mean separation of the points on the filament or sheet.

If we further assume that those idealised structures are mixed
randomly to form the point distribution we wish to study, the
Minkowski functionals $v_\mu$ are given by
Equation~(\ref{eq:poisson}), where the $\lV_\mu$ are a weighted
average over the three types of idealised structure.  We can therefore
calculate the $v_\mu$ numerically, extract the $\lV_\mu$, and perform
a linear fit to obtain the percentage for each type of structure in
the mixture.  By using the quantities $1-\lV_\mu/V_\mu(B)$, which are
exactly zero for field galaxies, we obtain
\begin{equation}
1-\frac{\lV_\mu}{V_\mu(B)} = 
\cF \left(1-\frac{\lV_\mu^{(\text{fil})}}{V_\mu(B)}\right) +
\cP \left(1-\frac{\lV_\mu^{(\text{sheet})}}{V_\mu(B)}\right)
\label{eq:toy}
\end{equation}
for the toy model.  We fix the free parameters $\cF$ and $\cP$ by
minimizing the $\chi^2$ of all four Minkowski functional profiles
at the same time.  This procedure yields two numbers for $\cF$ and
$\cP$; we interpret them as measures of filamentarity and planarity in
the point distribution.

The standard method of linear fitting solves the normal equations
{\cite{press:recipes}}, and also gives an error for our measurements,
usually of the order of 0.01.  As an example, Figure~\ref{fig:fit}
compares the Minkowski functionals of the CfA2N sample and the
best--fitting toy model.  Refinements of the fitting procedure, such
as a Principal Component Analysis {\cite{kendall:multivariate}} are
conceivable, but appear inappropriate in view of the vast
simplifications and resulting systematic problems of the toy model.
The assumption that galaxies arrange in infinitely thin filaments and
sheets is of course a strong simplification.  Nevertheless, this
simplified model already allows us to put a quantitative measurement
on our intuitive impression of the local Universe.

\end{document}